# Improving GPS Precision and Processing Time using Parallel and Reduced-Length Wiener Filters

Javier Garcia and Chi Zhou

**Abstract**— Increasing GPS precision at low cost has always been a challenge for the manufacturers of the GPS receivers. This paper proposes the use of a Wiener filter for increasing precision in substitution of traditional GPS/INS fusion systems, which require expensive inertial systems. In this paper, we first implement and compare three GPS signal processing schemes: a Kalman filter, a neural network and a Wiener filter and compare them in terms of precision and the processing time. To further reduce the processing time of Wiener filter, we propose parallel and reduced-length implementations. Finally, we calculate the sampling frequency that would be required in every Wiener scheme in order to obtain the same total processing time as the Kalman filter and the neural network.

———————————— ◆ ————————————

## 1 INTRODUCTION

GPS (Global Positioning System) is a global navigation satellite system that determines the position of any target by measuring the propagation delay of the signals from the satellites to the GPS receiver. Typically four or more satellites need to be tracked to calculate the position. GPS was developed by the US Department of Defense (DoD) in the 70s only for military purposes (positioning, navigation and weapons aiming). In fact, DoD included a distortion in the GPS signal called Selective Availability (SA) so that other people would not obtain good precision. Nowadays, however, GPS is used worldwide for civilian applications (e.g. driving assistance, topography, atmosphere study…) and that is why DoD decided to remove the SA.

Part of the current research on GPS is focused on increasing precision and combating signal outages by means of external aids, such as INS (Inertial Navigation Systems). The main problem here is how to combine the GPS signal with these aids. The two main solutions are using a Kalman filter and neural networks. The Kalman filter is a recursive filter used to estimate the state (in our case position) of a dynamic system and was proposed in 1960 by Rudolf Kalman [1] as an improvement over the linear predictors of that time. On the other hand, neural networks consist of interconnecting programming structures (neurons) that simulate the properties of the biological neural system. They adapt their response according to their training inputs so that they can be programmed for predicting the value of a signal considering different inputs (i.e. GPS and INS).

Another branch of the current research focuses on increasing precision just by means of processing the GPS signal (without INS). As [2] and [3] show, very accurate results can be obtained without spending money on installing inertial systems in every receiver. Again, the most frequently used mechanisms for this technique are Kalman filter and neural networks (and their different configurations: radial and fuzzy). In [2] Kalman filter and neural network performances are compared in a differential GPS scenario. The result is that the former is slightly more accurate but much slower. Therefore, if we want to keep the Kalman filter precision but also a fast response, we need to find a way of reducing its processing time. In [4] the authors propose using a parallel structure so that the filtering process is done concurrently. Unfortunately, the results are not very promising: processing time is reduced only by a 5.66% (from 6.36ms to 6ms). The main problem for implementing a parallel version of the Kalman filter is that it is not a linear filter (i.e. it has no impulse response), so there are not systematic means for achieving the parallel scheme.

Another possibility is using linear filters for prediction such as the Wiener filter. The Wiener filter was studied in [1] and proved to be a very accurate predictor filter. It is linear (so it has impulse response and, therefore, can be implemented in parallel) and optimum in terms of minimizing the MSE (mean squared error). At this point the Wiener filter seems very attractive but, why nobody has considered this filter for GPS? In fact, the Kalman filter was developed as an improvement over the Wiener filter [1]. Reference [5] also discusses the superiority of Kalaman filter over Wiener filter. The Wiener filter needs to accumulate past measurements in order to make a prediction, while the Kalman filter just needs the current measurement and the current state (updated in real time every sampling period); therefore, Kalman filter is usually preferred in real time applications [5].

————————————————

- *J. Garcia and C. Zhou are with the Electrical and Computer Engineering department, Illinois Institute of Technology, Chicago, IL.*

However, if the sampling rate was increased, the Wiener filter would accumulate the measurements so fast that users would feel that the prediction is performed in real time. The current GPS sampling rate is only 20Hz, which is understandable since GPS was developed in the 1970s. Therefore, if we increased the sampling rate, we would be able to apply other techniques (e.g. Wiener filter) to improve GPS precision even more. Furthermore, we can adjust the length of the Wiener filter (i.e. how many previous measurements we consider) in order to reduce the processing time, although we will get worse precision. So, there is a trade-off between processing time and precision.

The main goal of this work is two-fold. We first compare the Kalman filter, neural network and the Wiener filter in terms of the precision and prediction time for GPS applications. We then propose parallel and reduced-length implementations of the Wiener filter in order to improve the prediction time.

The paper is organized as follows. Section 2 explains MATLAB implementations of three different schemes: Kalman filter, neural network and Wiener filter and also presents their accuracy and processing time. Section 3 describes the implementation of the parallel and reduced-length versions of the Wiener filter, which are used to reduce the processing time. Section 4 calculates which sampling frequency would be required in every Wiener scheme in order to obtain the same total processing time as the Kalman filter and the neural network.

The simulations results show that the original Wiener filter is the most accurate but the slowest, while the neural network is the least accurate but the fastest; the Kalman filter is intermediate in both parameters. The improvement suggested in this work consists of either reducing the length of the Wiener filter or implementing it in parallel. Reducing the length of the filter decreases the precision (not much) but reduces the processing time significantly while implementing the filter in parallel keeps the original precision and reduces the processing time at the cost of using more hardware. Finally, increasing the sampling rate up to 30 KHz (which is a reasonable value for a sampling frequency nowadays), allows us to use most of the suggested variations for the Wiener filter.

## 2 PREDICTION FILTERS FOR GPS APPLICATIONS

We use the "Constellation Toolbox for MATLAB" by Constell, INC to obtain position vectors for GPS and then use it to implement each of the prediction filters explained in the sequel. The Constellation Toolbox is an integrated collection of MATLAB files that allows the user to model, simulate and analyze satellite constellations. It provides modeling capabilities for the GPS and GLONASS constellations for a great variety of navigation applications. We will only focus on the position vectors of the GPS receiver. These position vectors are contained in 'alamanac' file in a matrix form where any row contains a position vector and the next row contains the position vector for 50 ms later (since the sampling frequency is equal to 20 Hz). These vectors have a length of 181 but in this work it has been limited to 180. In the following the 'state' of the filter corresponds to the position vector.

### 2.1 The Kalman Filter

The Kalman filter solves the problem of estimating the state $x \in \Re^n$ of a time controlled process directed by the linear stochastic difference equation:

$$x_k = A \cdot x_{k-1} \quad (1)$$

$A$ is an *n times n* matrix (where $n$ is the number of state variables considered by the Kalman Filter) that relates the state at the previous time *k-1* to the current state at time *k*. $A$ may change at every time step.

We employ the following notation: $x_k$ represents the process state vector at time k, $\hat{x}_k^-$ represents the a priori state estimate (estimation of $x_k$ before updating with the current measurement $z_k$) and $\hat{x}_k$ represents the a posteriori state estimate (estimation of $x_k$ after updating with the current measurement $z_k$).

The a posteriori estimate $\hat{x}_k$ is computed as a linear combination of the a priori estimate $\hat{x}_k^-$ and a weighted difference between the current measurement $z_k$ and a measurement prediction $H$:

$$\hat{x} = \hat{x}_k^- + K_k(z_k - H\hat{x}_k^-) \quad (2)$$

where $H$ is an *mxn* matrix (where $m$ is the number of measurements that we take) that relates the state to the measurement $z_k$; which may change with each time step. The difference $(z_k - H\hat{x}_k^-)$ is called residual and it reflects the discrepancy between the predicted measurement $H\hat{x}_k^-$ and the current measurement $z_k$. $K_k$ is an nxm matrix called gain that aims to minimize *a posteriori* error covariance. It can be calculated as:

$$K_k = P_k^- \cdot H \cdot (HP_k^- H^T + R)^{-1} \quad (3)$$

where $R$ is the covariance of the error. Generally it is easy to determine because we also need to measure the process so we should be able to obtain some off-line sample measurements. As $R$ decreases, the current measurement $z_k$ is trusted more in the second equation, while the predicted measurement $H\hat{x}_k^-$ is trusted less. $P_k$ is the estimate of the error covariance matrix; it is calculated as:

$$P_k^- = A \cdot P_{k-1} \cdot A^T + Q \quad (4)$$

where $Q$ is the noise covariance. The noise covariance is typically difficult to determine since we do not usually have the chance to directly observe the process we are estimating. If both $Q$ and $R$ are constant, both the estimation error covariance $P_k$ and the Kalman gain $K_k$ will converge quickly.

The error covariance matrix $P_k$ is updated every time step by the following equation:

$$P_k = (1 - K_k H) \cdot P_k^- \quad (5)$$

The filter goes making estimations periodically and then obtains feedback from the noisy measurements; therefore, the equations for the Kalman filter are classified into two groups: time update equations and measurement update equations. The former ones use the current state and the error covariance estimates to obtain the a priori estimates; the latter ones incorporate the new measure-ment into the a priori estimate to produce the a posteriori estimate. We have a summary of the structure of Kalman filter in Fig. 1.

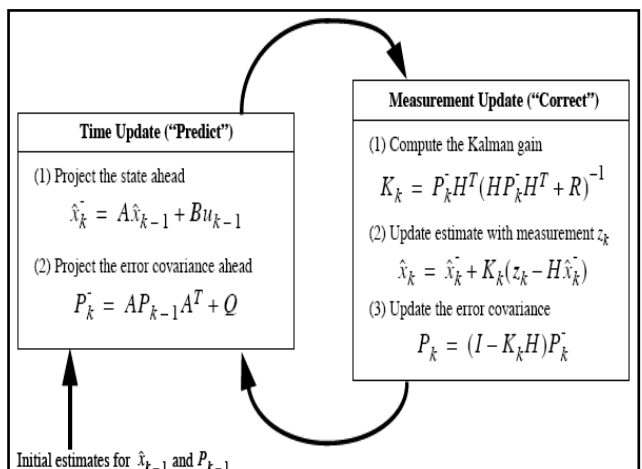

Fig. 1. The original Kalman filter equations. [5]

The steps involved in implementing the Kalman filter for the z component of the GPS signal (called v_z) are as follows:
- We set $n = 2$ because two state variables are considered: position and velocity.
- $m = 1$ because we will work with one measurement at each time step.
- The estimation for the initial position is the mean of the two first samples, while the estimation for the initial velocity is the difference of the two first samples divided by the sampling period (50 ms).
- The initial a priori estimate of the error covariance matrix $P_0^-$ is equal to the noise covariance $Q$ multiplied by the identity matrix of order $n=2$. It is difficult to measure the value of $Q$, so the chosen value is $Q = 30^2$ as in [6].
- The measurement matrix chosen is H = [1, 0] because in GPS we only receive position measurements and not velocity ones.
- The state update matrix A is:

$$A = \begin{bmatrix} 1 & 0.05 \\ 0 & 1 \end{bmatrix} \quad (6)$$

since the new measurement is the old one and the current velocity multiplied by the sampling period (50 ms). Below is a summary of equations:

$$\begin{cases} X_0^- = \begin{bmatrix} Position_z \\ Velocity_z \end{bmatrix} = \begin{bmatrix} \dfrac{v_{z(1)} + v_{z(2)}}{2} \\ \dfrac{v_{z(2)} + v_{z(1)}}{50ms} \end{bmatrix} \\ P_0^- = Q \cdot \begin{bmatrix} 1 & 0 \\ 0 & 1 \end{bmatrix} = 30^2 \cdot \begin{bmatrix} 1 & 0 \\ 0 & 1 \end{bmatrix} \\ H = \begin{bmatrix} 1 & 0 \end{bmatrix} \end{cases} \quad (7)$$

## 2.2 Neural Networks

A neural network constitutes a mathematical model that implements a function $f(x)$, which is the result of composing other functions $g_i(x)$, which are themselves the result of composing other functions $h_i(x)$. These functions are usually represented as a network structure with arrows showing the relations between variables. The most frequently used composed function is the so-called "non-linear weighted sum", which is defined as:

$$f(x) = K \cdot \left( \sum_i \omega_i \cdot g_i(x) \right) \quad (8)$$

where $K$ is some known function (e.g. the hyperbolic tangent). It is also usual to refer to a collection of functions $g_i$ as simply a vector $g = (g_1, g_2, \ldots, g_n)$.

The main improvement of the neural networks with respect to Kalman filters is that there are adaptive training criterions in which neurons learn how to face new inputs. Learning consists of using a set of measurements to find a new function f', so that we can derive a cost function C such that, there is no solution that has a cost less than the cost of the optimal solution f'. The cost function C determines the discrepancy of the current solution from an optimal solution; learning implies searching through the solution space for a function that has the smallest possible cost.

The neural network implemented in this paper is a multi-layer perceptron feedforward one, which has 2 inputs, 3 hidden nodes and 1 output. A supervised learning algorithm called backpropagation [7] has been used for training the network; this algorithm does not require any feedback (so it is adequate for this feedforward network). In this algorithm we compute the gradient of the error with respect to the assigned weights; therefore errors propagate from the output nodes to the inner nodes. The computed gradient is employed to determine the weights that minimize the error. Backpropagation usually allows fast convergence on satisfactory local minima for error in the kind of networks to which it is suited.

The Backpropagation algorithm has already been implemented in MATLAB by KashaniPour and Phil Brierley. It consists of the following steps:
1. Uses a certain training sample to the network.

2. Compare the output obtained using this training sample to the desired output from that sample.
3. Determine the error in each output neuron.
4. Determine what output we should have obtained for each neuron, and also a scaling factor indicating how much the output must be adjusted to match the desired output. This is known as the local error.
5. Modify the weights of each neuron to reduce the local error of the last step.
6. Modify the neurons at the previous levels that are responsible for the local, increasing the responsibility to those neurons connected by stronger weights.

We modify the above existing algorithm so that the first 90 samples of the received GPS vector $v\_z$ have been used for training the neural network and then the error is calculated for the last 90 samples.

### 2.3 The Wiener Filter

This filter was developed in 1949 by Norbert Wiener in his book [8]. The objective of the Wiener filter is to remove the noise that has corrupted a signal by means of a statistical approach. It is required to have information about the spectral or the statistical properties (power spectral density or auto-correlation and cross-correlation) of the original signal and the noise, so that we can design a LTI filter whose output is as similar as the original signal as possible. The theory of Wiener filters is focused on those filters that are causal, that is, the ones that work only with the past and present of the time series. In this filter, the criterion selected for optimizing the filter impulse response is the minimization of the MSE.

$$\begin{bmatrix} r_{xx}(0) & r_{xx}(1) & \cdots & r_{xx}(M-1) \\ r_{xx}(1) & r_{xx}(0) & \cdots & \vdots \\ \vdots & \vdots & & r_{xx}(1) \\ r_{xx}(M-1) & & & r_{xx}(0) \end{bmatrix} \cdot \begin{bmatrix} h(0) \\ h(1) \\ \vdots \\ h(M-1) \end{bmatrix} = \begin{bmatrix} r_{sx}(0) \\ r_{sx}(1) \\ \vdots \\ r_{sx}(M-1) \end{bmatrix}$$

$$\Rightarrow [R_{xx}] \cdot [h] = [r_{sx}] \Rightarrow [h] = [R_{xx}]^{-1} \cdot [r_{sx}] \qquad (10)$$

We will consider that the input of the Wiener filter is a signal $s(t)$ corrupted by additive noise $n(t)$. The output $\hat{s}(t)$ is calculated by filtering the input signal with the Wiener filter impulse response $g(t)$.

$$\hat{s}(t) = g(t) * (s(t) + n(t)) \qquad (9)$$

After some calculations [8], we obtain the expression of the FIR filter coefficients:

where:
- $M$ is the desired length of the filter
- $r_{xx}(n)$ is the autocorrelation function of the noisy signal
- $r_{sx}(n)$ is the crosscorrelation function between the signal and the noise.

### 2.4 Comparison Study

1. Simulation Results:

The result achieved after applying Kalman Filtering is shown in Fig. 2.

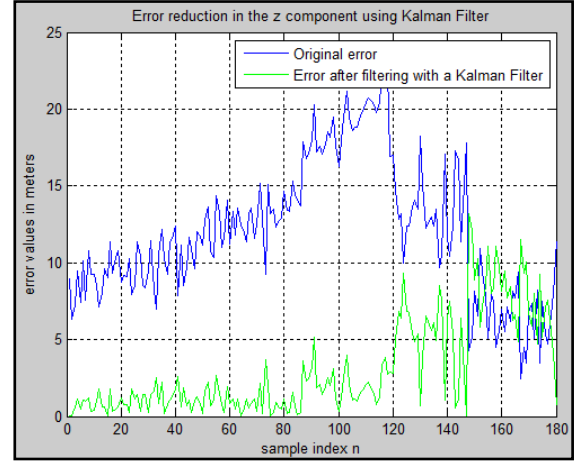

Fig. 2. Error after applying the Kalman filter.

The errors after applying Kalman filter are calculated as the difference (in absolute value) between the output of the Kalman filter and the constant signal. Here we observe that the Kalman filter reduces the error considerably (mean = 3.0971 meters, variance = 9.7733 meters). However, when there is a sudden change in precision (e.g. sample 118 and sample 148) the Kalman filter increases the error, even if the sudden change has reduced the error in the original signal.

The error obtained after applying the neural network can be seen in Figure 3.

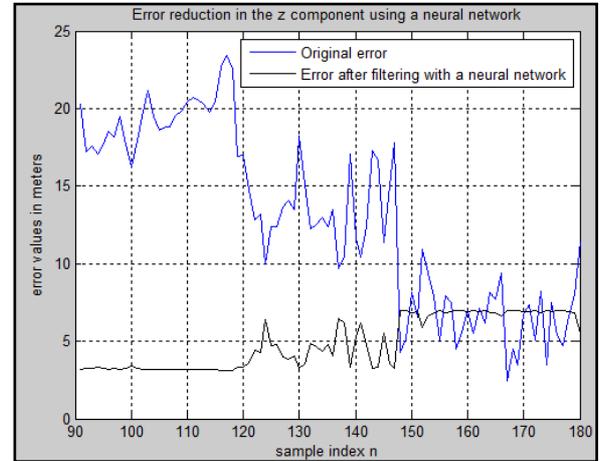

Fig. 3. Error after applying the neural network

For the simulation result in Fig. 3, we trained the neural network using the first 90 samples, so we can only evaluate its performance for the rest of the samples (from the 91st to the 180th). We observe that the neural network also improves the original error significantly (mean = 4.7119, variance = 3.0141). The mean of the error is larger than the one for the Kalman filter, but the error is more constant (i.e. the variance is smaller); this is due to the main property of the backpropagation algorithm. Backpropagation allows fast convergence on satisfactory local minima for error in the kind of networks to which it is suited; that is, the error in the neural network is quite constant as long as the original error does not vary too much, however, when there is a peak in the original error, the error in the neural networks start to fluctuate.

Also the error obtained after applying the Wiener filter can be seen in Fig. 4

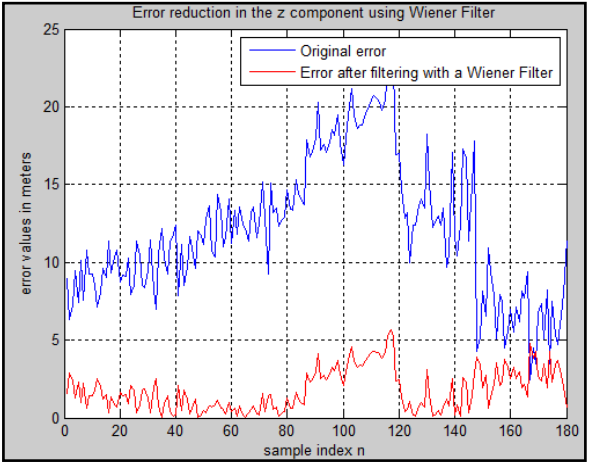

Fig. 4. Error after applying the original Wiener filter

We see that the Wiener filter achieves the smallest error in both mean and average (mean = 1.7938, variance = 1.7804). However, as we will analyze later, the Wiener filter is the slowest and also it needs to accumulate the 180 samples for calculating the coefficients of the filter.

2  Precision comparison:

Table I shows the error (in absolute value) for each scheme. It is calculated as the difference between the real position vector and the output of each scheme.

TABLE I: Precision comparison

|  | Original | Kalman | Neural | Wiener |
|---|---|---|---|---|
| Mean(m) | 12.0916 | 3.0971 | 4.7119 | 1.7938 |
| Variance | 20.058 | 9.7733 | 3.0141 | 1.7804 |

It can be seen from the above table that the Kalman filter reduces the mean of the error enormously (from 12 meters to 3), but its variance is still quite large. The neural network is slightly worse, but its error is more constant since the Backpropagation algorithm allows fast convergence. Finally, the Wiener filter is the most precise in both mean and variance. Neural filter requires 80 samples and wiener filter requires 180 samples in the above table for calculating the coefficients of the filter.

3  Processing time comparison:

Table II shows the processing time for each scheme. It does not include the acquisition time which has to be summed in the Wiener filter case.

TABLE II: Processing time comparison

|  | Kalman | Neural | Wiener |
|---|---|---|---|
| Processing time (ms) | 20.3281 | 16.9648 | 25.2656 |

Table II indicates that the neural network is the fastest scheme and the original Wiener filter is the slowest one; that is why two methods for reducing the processing time are developed in this work: the parallel implementation and the reduced length one.

## 3 PARALLEL AND REDUCED-LENGTH ARCHITECTURES

### 3.1 The Parallel Acrchitecture

Now that we have implemented the Wiener filter, we can try reducing the processing time. A reasonable way of doing this is performing the filtering process in parallel. The main inconvenient for this approach is that it requires incrementing the hardware, but we only need to add filters, delayers and adders which are very cheap and will not make the GPS receiver much more expensive. Reference [9] explains how to implement the 2-parallel and the 3-parallel structures in an efficient way. Figure 5 shows how to implement the 2-parallel scheme.

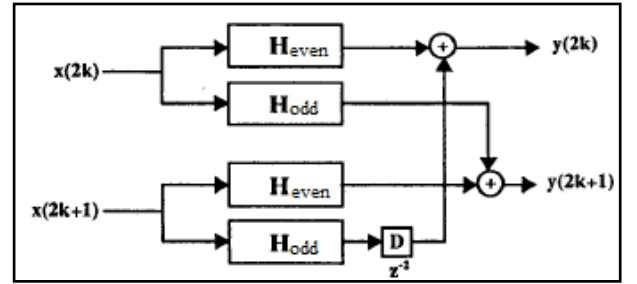

Fig. 5. The 2-parallel implementation [9].

The even and odd outputs are calculated as:

$$\begin{cases} Y_{even}(z^2) = H_{even}(z^2) \cdot X_{even}(z^2) + z^{-2} \cdot H_{odd}(z^2) \cdot X_{odd}(z^2) \\ Y_{odd}(z^2) = H_{odd}(z^2) \cdot X_{even}(z^2) + H_{even}(z^2) \cdot X_{odd}(z^2) \end{cases} \quad (11)$$

Finally the total output is expressed as:

$$Y(z) = Y_{even}(z^2) + z^{-1} \cdot Y_{odd}(z^2) \quad (12)$$

Fig. 6 shows the 3-parallel implementation.

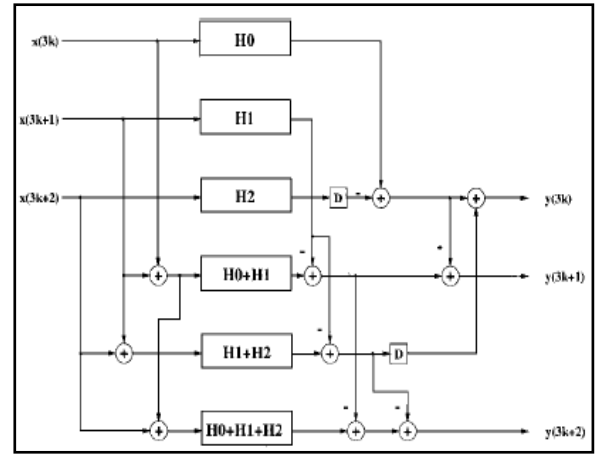

Fig. 6. The 3-parallel implementation [9].

Here:

$$\begin{cases} H_0(z^3) = \{h(0),0,0,h(3),0,0,h(6),\ldots\} \\ H_1(z^3) = \{h(1),0,0,h(4),0,0,h(7),\ldots\} \\ H_2(z^3) = \{h(2),0,0,h(5),0,0,h(8),\ldots\} \end{cases} \quad (13)$$

The partial outputs here are calculated as:

$$Y_0 = H_0 X_0 - z^{-3} H_2 X_2 + z^{-3}[(H_1+H_2)(X_1+X_2) - H_1 X_1] \quad (14)$$

$$Y_1 = [(H_0+H_1)(X_0+X_1) - H_1 X_1] - [H_0 X_0 - z^{-3} H_2 X_2] \quad (15)$$

$$Y_2 = [(H_0+H_1+H_2)(X_0+X_1+X_2)] - \\ [(H_0+H_1)(X_0+X_1) - H_1 X_1] - [(H_1+H_2)(X_1+X_2) - H_1 X_1] \quad (16)$$

Finally the total output is:
$$Y(z) = Y_0(z^3) + z^{-1} \cdot Y_1(z^3) + z^{-2} \cdot Y_2(z^3) \quad (17)$$

### 3.2 The Reduced-length Wiener Filter

This is another approach for reducing the processing time: instead of waiting for receiving the 180 samples and then performing 180 samples-convolutions (as in the original simulation), we reduce the length of the Wiener filter to $x$, so that we have to wait for receiving only x samples and then perform $x$ samples-convolutions. This approach has the advantage that it does not require more hardware. In Figs. 7 and 8, we will compare the 135-samples and the 90- samples versions via simulations.

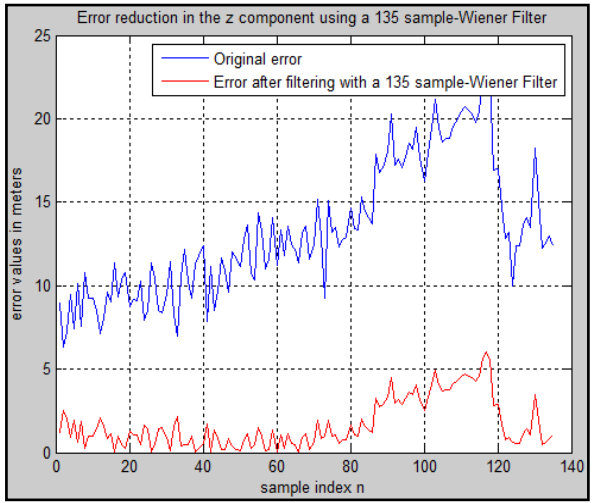

Fig. 7. Error after applying the 135 sample Wiener filter

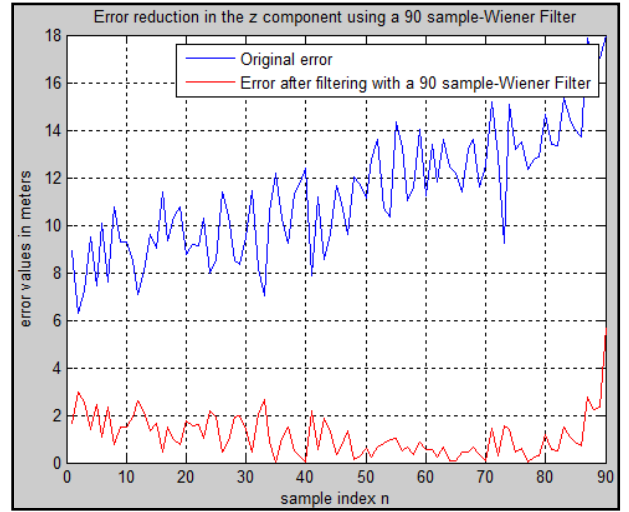

Fig. 8. Error after applying the 90 sample-Wiener filter

Finally, we plot the last 90 samples of the Kalman filter, the neural network and the original Wiener filter together in Fig. 9.

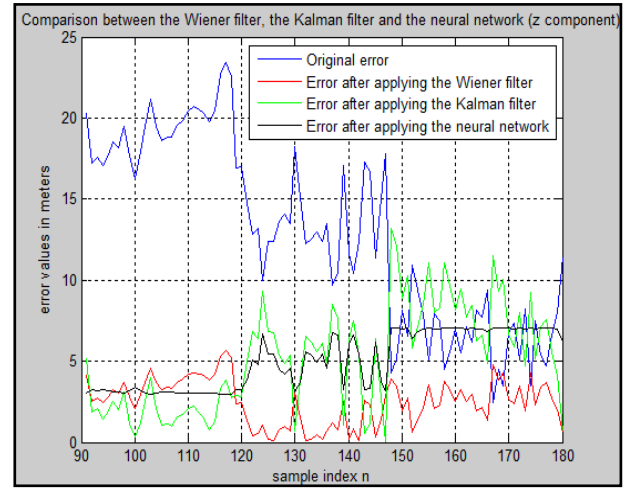

Fig. 9. Comparison between the three techniques

### 3.4 Proposed Wiener Filter Architecture Comparisons

Here we compare the precision and processing time of the proposed parallel and reduced-length architectures for Wiener filter. We can see the results on Table III:

TABLE III: Parallel and reduced-length architectures comparison for Wiener filter

|  | Original | //2 | //3 | L=135 | L= 90 |
|---|---|---|---|---|---|
| Mean(m) | 1.7938 | 1.7938 | 1.7938 | 2.3545 | 3.102 |
| Variance | 1.7938 | 1.7938 | 1.7938 | 4.316 | 8.287 |
| Processing | 25.2656 | 20.234 | 15.1483 | 15.5938 | 9.688 |

We observe that the parallel implementation does not degrade the original Wiener filter accuracy; however, the

97processing time is reduced by a small amount. The reduced length architecture improves the processing time substantially in exchange for some degradation of precision.

## 4 SAMPLING FREQUENCY REUIREMENTS

In this section we compare the total prediction time. Since both the Kalman filter and the neural network are real-time mechanisms (i.e. they work with the current measurement so they do not need to accumulate samples), the processing time is equal to the total time for them. The current sampling frequency in GPS (which has not changed since its beginning) is only 20 Hz, so its sampling period is 50 ms, which is already larger than any processing time; this way, accumulating 180 samples in a Wiener filter would take 9 seconds, which is incompatible with any real-time application. However, the value of 20 Hz for a sampling frequency is nowadays extremely small. Increasing this frequency to some KHz would allow us to use the Wiener filter in real-time applications. Now we are going to calculate the required sampling frequency for getting the same processing time as the Kalman filter and the neural network.

The processing time of the original Wiener filter is already larger than that of both the Kalman filter and the neural network, so there is no way to equalize their total times. The processing time of the 2-parallel Wiener filter is already larger than that of the neural network, but we can equalize the total time of the neural network. Let us calculate which sampling frequency we would need:

$$T_{kalman} = T_{//2} + 180\,samples \cdot T_{sampling-//2} \Rightarrow$$
$$f_{sampling,//2} = \frac{1}{T_{sampling,//2}} = \frac{1}{(T_{kalman} - T_{//2})/180} = 1912.86\,KHz$$

This sampling frequency is too high, so we should switch to another configuration. The total time of the reduced length configurations is considerably smaller because, apart from having a smaller processing time, they have to accumulate fewer samples. We calculate the reuired sampling frequency for the 90-samples Wiener filter to equalize the total time of the neural network as:

$$T_{neural} = T_{90} + 90\,samples \cdot T_{sampling-90} \Rightarrow$$
$$f_{sampling-90} = \frac{1}{T_{sampling-90}} = \frac{1}{(T_{neural} - T_{90})/90} = 12.37\,KHz$$

This sampling frequency is much more reasonable. Table IV shows all the different combinations:

TABLE IV: Sampling frequencies needed in KHz for parallel and reduced-length architectures of Wiener filter

|        | Original   | //2        | //3   | L= 135 | L= 90 |
|--------|------------|------------|-------|--------|-------|
| Kalman | Impossible | 1912.86    | 34.75 | 28.52  | 8.46  |
| Neural | Impossible | Impossible | 99.09 | 98.47  | 12.37 |

As we see in this table, 35 KHz is enough for most configurations to equalize the total time of the Kalman filter, and 100 KHz is enough for them to equalize the total time of the neural network.

## 4 CONCLUSION

In this paper we have studied three different techniques for improving precision at the GPS receiver without need for INS information. They are: Kalman filter, neural network and Wiener filter. The simulations show that the Wiener filter is the most accurate method. However, the main disadvantage of the Wiener filter is that it requires additional time for accumulating measurements for determining the filter coefficients. Therefore we should increase the GPS sampling frequency for accumulating these measurements faster. This work shows that increasing the sampling frequency to 35 KHz lets us obtain similar total processing time to the Kalman filter (but better precision) in most of the proposed configurations; whereas 100 KHz is enough to equalize neural networks time.

Two schemes are proposed for reducing the processing time of the Wiener filter: Implementing the filter in parallel and reducing its length. The former requires increasing the hardware (i.e. more filters, adders and delayers) and gets exactly the same accuracy as the original filter since they are analytically equivalent; the latter does not need any additional hardware and reduces the processing time significantly, but it makes precision slightly worse.

A possible improvement to this work consists of combining the two proposed schemes: reducing the length of the filter and then parallelizing its implementation. Using this improvement method, we should get the same accuracy as the reduced-length implementation, but a shorter processing time. Another possibility is implementing an IIR Wiener filter instead of a FIR one. The IIR is optimum in terms of minimizing the MSE, but we have to make sure that our implementation is stable and causal; otherwise it will not be applicable in GPS applications.